\documentclass[12pt]{iopart}

\usepackage{graphicx}
\usepackage{bm}
\usepackage[normalem]{ulem}
\usepackage[usenames,dvipsnames]{color}

\usepackage{amssymb}

\newcommand{\Xlabel}{''}
\newcommand{\Alabel}{'}
\newcommand{\Clabel}{'''}

\newcommand{\moleculenumber}{$6 \times 10^4$\,}

\newcommand{\myfirst}{1$^\textrm{\footnotesize st}$\ }
\newcommand{\mysecond}{2$^\textrm{\footnotesize nd}$\ }

\newcommand{\rfig}[1]{Fig.~\ref{#1}}

\makeatother

\usepackage{cite} 

\usepackage{hyperref}
\hypersetup{colorlinks, citecolor=Blue, linkcolor=Blue, urlcolor=Blue}
\begin{document}

\title[]{Laser slowing of CaF molecules to near the capture velocity of a
molecular MOT}

\author{Boerge Hemmerling$^{1,2,7,*}$, Eunmi Chae$^{1,2,*}$,
   Aakash Ravi$^{1,2}$, 
   Loic Anderegg$^{1,2}$,
   Garrett K.~Drayna$^{2,3}$, 
   Nicholas R.~Hutzler$^{1,2}$,
   Alejandra L.~Collopy$^{5,6}$,
   Jun Ye$^{5,6}$, 
   Wolfgang Ketterle$^{2,4}$,
   and John M.~Doyle$^{1,2}$
}

\address{$^1$Department of Physics, Harvard University, Cambridge, Massachusetts
02138, USA}

\address{$^2$Harvard-MIT Center for Ultracold Atoms, Cambridge, Massachusetts
02138, USA}

\address{$^3$Department of Chemistry and Chemical Biology, Harvard University, Cambridge, Massachusetts
02138, USA}

\address{$^4$Department of Physics, Massachusetts Institute of Technology, Cambridge,
Massachusetts 02139, USA}

\address{$^5$JILA, National Institute of Standards and Technology and University
of Colorado, Boulder, Colorado 80309, USA}

\address{$^6$Department of Physics, University of Colorado, Boulder, Colorado
80309, USA}

\address{$^7${\it Present address:} Department of Physics, University of California, Berkeley, California 94720, USA}

\address{$^*$These authors contributed equally to this work.}

\ead{eunmi@cua.harvard.edu}

\begin{abstract}
   Laser slowing of CaF molecules down to the capture velocity of a magneto-optical trap (MOT) for molecules is achieved. Starting from a two-stage buffer gas beam source, we apply frequency-broadened \textquotedblleft white-light\textquotedblright{} slowing and observe approximately \moleculenumber CaF molecules with velocities near 10\,m/s. CaF is a candidate for collisional studies in the mK regime. This work represents a significant step towards magneto-optical trapping of CaF.
\end{abstract}


\maketitle


The creation and control of samples of ultracold atoms enabled many milestones in atomic physics, such as Bose-Einstein condensation \cite{Davis1995,Anderson1995}, quantum simulations of many-body systems \cite{Bloch2008,Bloch2012}, development of quantum information systems \cite{Saffman2010}, precision measurements and atomic clocks \cite{Cronin2009,Ludlow2015}. Ultracold polar molecules may advance these and other areas of science even further, owing to the molecules' additional degrees of freedom, large electric dipole moments, and chemical characteristics. These properties are at the core of many proposals and experiments \cite{Carr2009}, including quantum simulation of strongly correlated systems \cite{Baranov2012,Buchler2007,Micheli2006}, precision measurements and tests of fundamental physics \cite{Baron2014,DeMille2008}, quantum information processing \cite{Andre2006,DeMille2002,Rabl2006}, studies of ultracold collisions \cite{Sawyer2011,Hummon2011}, and control of ultracold chemical reactions \cite{Schenll2009,Ni2010h,Ospelkaus2010}. However, a basic requirement for many proposed experiments is trapped molecules at temperatures around 1\,mK or below.

Coherent association of ultracold atoms has been successful in generating ultracold ground state molecules at high phase-space densities \cite{Ni2008,Park2015a,Takekoshi2014,Molony2014,Wang2016}. However, this approach so far is restricted to atomic species with easily accessible laser cooling transitions. This method is not yet applicable to a large variety of molecules, including free radicals such as calcium monofluoride (CaF). An alternative approach is the direct cooling of molecules. A direct cooling scheme typically starts with slowing of molecules to load a trap, where further cooling can take place to reach ultracold temperatures. 

Various approaches have been pursued for trapping molecules, including electrostatic traps \cite{Bethlem2000,VandeMeerakker2005h,Hoekstra2007e,Zeppenfeld2012a,Prehn2016}, magnetic traps \cite{Weinstein1998,Sawyer2007,Campbell2007,Hummon2011c,Lu2014,Stuhl2012e,Stoll2008,Riedel2011}, and magneto-optical traps (MOTs) \cite{Hummon2013,Barry2014,McCarron2015,Norrgard2015}. These traps are typically $\lesssim 1$\,K deep, and therefore a source of cold molecules is necessary.  The highest intensity source of cold and slow molecules is the buffer gas beam \cite{Maxwell2005,Patterson2007,Lu2011,Hutzler2012}, which utilizes collisions with an inert, cryogenic gas. However, even these slow sources still typically inhibit direct loading of traps like a MOT since the vast majority of molecules have velocities above the trap's capture velocity. Hence, an initial slowing stage is required to provide a significant fraction of molecules that can be trapped. At present, several groups have managed to slow molecules in various ways \cite{Bethlem2000,Riedel2011,Narevicius2012,Akerman2015,Chervenkov2014}, including laser cooling and radiation pressure slowing \cite{Barry2012a,Zhelyazkova2014,Yeo2015a}. Here, we present laser slowing of CaF molecules, originating from a two-stage cryogenic buffer-gas beam source \cite{Patterson2007,Lu2011}, to velocities around $10$\,m/s, which is near the expected capture velocity of a molecular MOT for CaF.

The complex internal structure of molecules renders the use of a Zeeman slower difficult. Instead, \textquotedblleft white-light\textquotedblright{} slowing is used, in which spectrally broadened lasers counter-propagate with respect to the molecular beam to address a range of velocity classes and the internal hyperfine structure of the molecules, as the molecules decelerate \cite{Barry2012a,Yeo2015a}. Due to the divergence of the molecular beam, a significant fraction of the molecules do not reach the MOT capture volume. As a result, the total number of molecules inside a molecular MOT has never surpassed 2000 in recent experiments \cite{Barry2014,McCarron2015,Norrgard2015}. To realize many of the proposed applications of trapped molecules, such as evaporation of the trapped sample in a subsequent (magnetic) trap, higher molecule numbers are required. The most straightforward way to increase the number of molecules that reach the MOT capture volume is to shorten the distance required for slowing, which in turn increases the solid angle of molecules captured from the source. This is achieved here by starting from slower initial forward velocities by using a two-stage cryogenic buffer-gas beam (CBGB) source and by using a low mass molecule (which can be decelerated over a shorter distance), at the cost of lower on-axis beam intensity compared to a single stage source. This two-stage CBGB produces a molecular beam with peak forward velocity of 60\,m/s, more than a factor of two slower than a single-stage buffer-gas beam source, and has a flux of $\sim 10^9$ molecules/steradian/state/pulse \cite{Lu2011}. 

\begin{figure*}
\begin{centering}
   \includegraphics[scale=0.5]{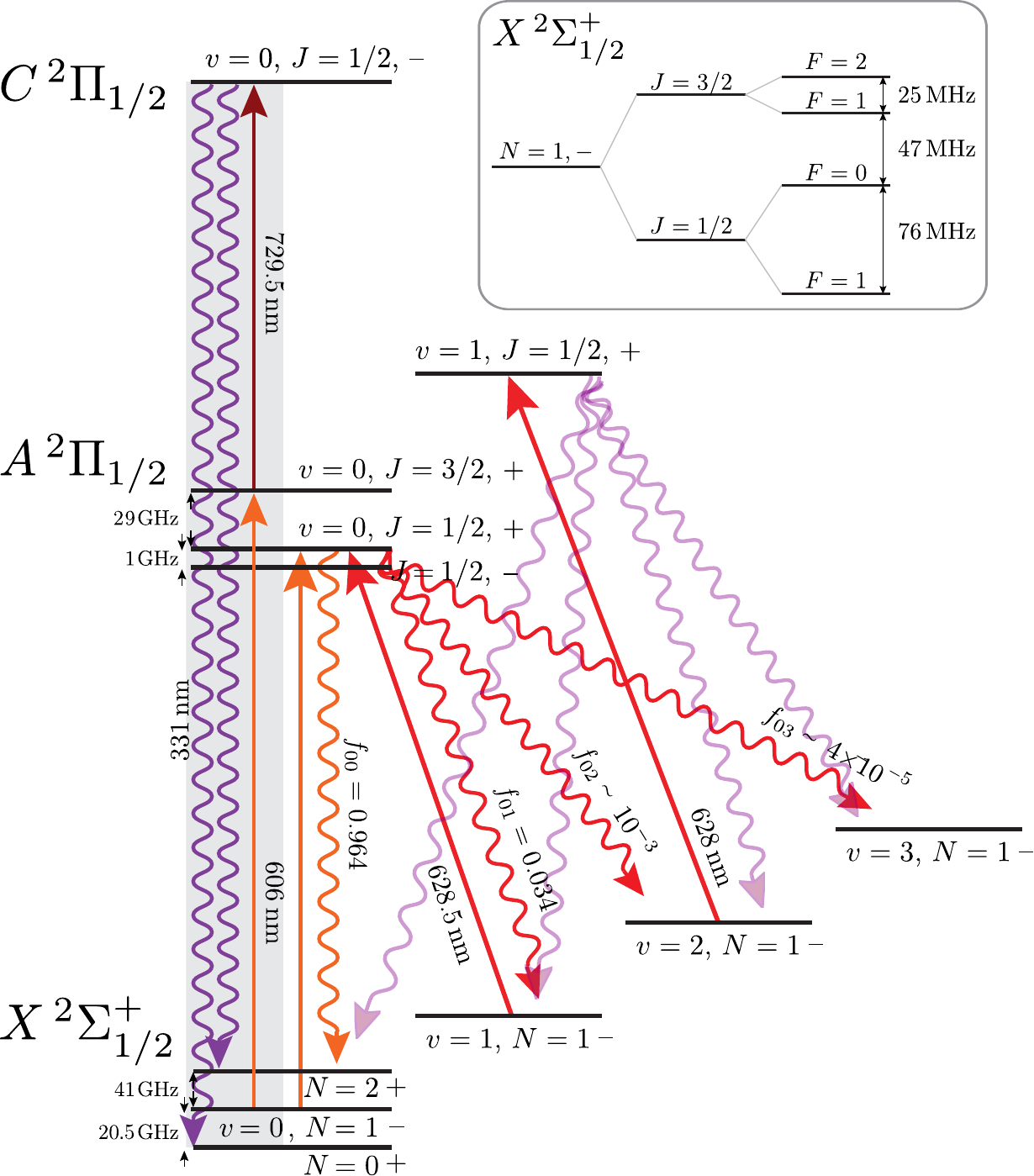}
\par\end{centering}

\protect\caption{\label{fig:CaFEnergyLevel}Relevant energy levels of a CaF molecule. $v$ and $N^p$ denote the vibrational and rotational quantum number respectively, where $p=+/-$ denotes the parity of the state. $f_{ij}$ represents the Franck-Condon factor between the $X\,(v\Xlabel=j)$ and $A\,(v\Alabel=i)$ states from theoretical values in Ref. \protect\cite{Pelegrini2005}. The value of $f_{00}$ has been measured to be 0.987 \protect\cite{Wall2008}. Straight lines indicate applied laser frequencies and wavy lines indicate spontaneous emission. Transitions in the gray box are used for detecting molecules (see text). The spin-rotation and hyperfine interactions of the electronic ground state $X$ are depicted in the inset. The hyperfine structure in the excited states is not resolved.}
\end{figure*}

CaF is a favorable candidate for laser cooling due to its highly diagonal Franck-Condon factors (with a measured $A\,(v\Alabel=0) - X(v\Xlabel=0)$ branching ratio of $f_{00}$ = 0.987) \cite{Wall2008,Pelegrini2005}. It is a $^{2}\Sigma$ molecule which has a free electron in its outermost orbital. This electron\textquoteright s spin degree of freedom makes CaF attractive and distinct from bi-alkali molecules. It also has a large electric dipole moment of 3\,Debye. The relevant energy levels of CaF are shown in \rfig{fig:CaFEnergyLevel}. The lowest electronic excited state ($A\,^{2}\Pi_{1/2}$) has a lifetime of 19.2 ns \cite{Wall2008}.

We follow the laser cooling scheme reported in Ref. \cite{Barry2012a}. A CaF molecule can scatter about $10^{5}$ photons with the $X(v\Xlabel=0)-A(v\Alabel=0)$ (main) laser and the two vibrational repump lasers (\rfig{fig:CaFEnergyLevel}) before decaying into the higher vibrational states. Rotational branching within each vibrational manifold is avoided by driving a $P$(1) rotational transition \cite{Stuhl2008}. Due to the interaction of the electron spin $S = 1/2$ and the fluorine\textquoteright s nuclear spin $I = 1/2$, the rotational state splits into four hyperfine states. Each of these states need to be addressed with laser radiation to keep the molecules in the optical cycle (\rfig{fig:CaFEnergyLevel} inset). All slowing lasers are spectrally broadened to cover the hyperfine splittings in the ground states and to compensate the changing Doppler shift as the molecules are decelerated. The maximum starting velocity of a molecule that can be addressed in this configuration is then determined by the total width of the spectra of the slowing lasers and their interaction time with the molecules. While a broad spectrum allows for addressing high velocities, the resulting lower power density limits the scattering rate which, in turn, limits the deceleration rate. An alternative approach that would maintain a higher power density is chirped slowing, which was successfully implemented with YO \cite{Yeo2015a}. This latter approach was not employed by us as frequency chirping of the slowing laser pulses is limited by the finite bandwidth of the laser servos ($< 100$\,Hz).

\begin{figure*}
\begin{centering}
\includegraphics[width=4.5in]{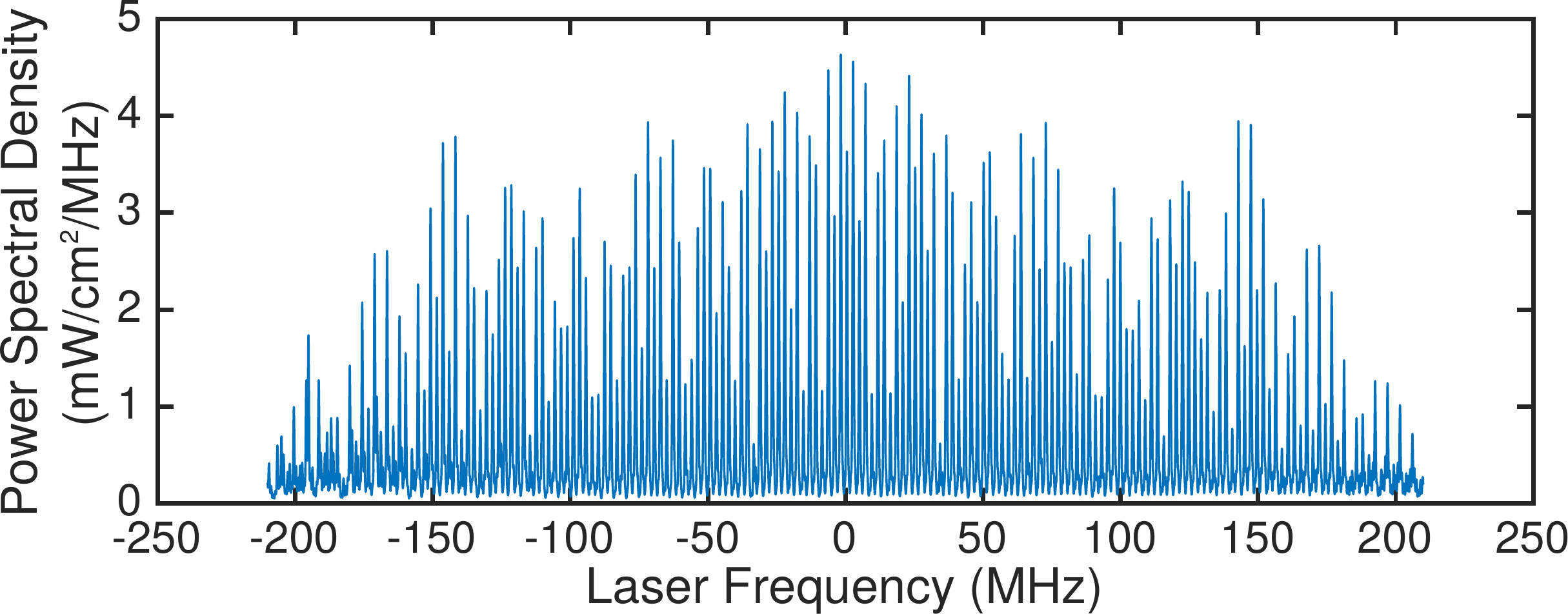}
\par\end{centering}
   
\protect\caption{\label{fig:MainBroadening}Spectrum of the broadened main slowing laser. The zero of the $x$-axis is set as the center of the broadened spectrum.
The first and the second repump lasers are broadened to approximately the same width.}
\end{figure*}

\rfig{fig:MainBroadening} shows the spectrum of the main slowing laser. The spectrum is broadened by two sequential phase-modulating electro-optic modulators (EOMs): One covers the hyperfine splitting of the ground state with a 24.8 MHz modulation frequency with a modulation index of $\approx 4$. The second EOM is driven at a frequency of 4.5 MHz with a modulation index of $\approx 17$. This drive frequency is chosen to be about half the excited state linewidth of 8.29 MHz to maintain the resonance condition throughout the slowing process. The total broadening spans about 400 MHz, which covers all the hyperfine splittings and the velocity change.

\begin{figure*}
\begin{centering}
\includegraphics[width=0.75\textwidth]{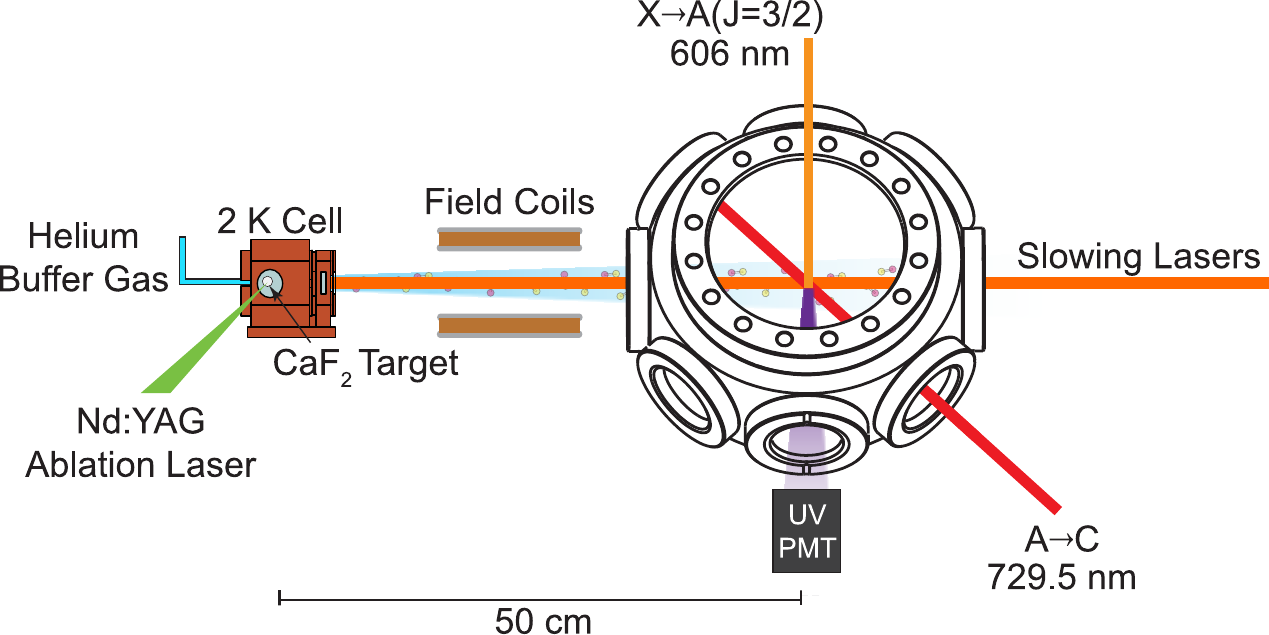}
\par\end{centering}

\protect\caption{\label{fig:ExperimentalSetup}Experimental Setup (not to scale). A two-stage cell at around 2 K produces a slow molecular beam with forward velocity of about 60 m/s. The $X\rightarrow A$ laser orthogonal to the molecular beam and the $A\rightarrow C$ laser at 45\,degrees relative to the molecular beam intersect at 50\,cm downstream where the molecules are detected. UV photons that are emitted when the molecules decay from the $C$ state to the $X$ state are detected by a photomultiplier tube. The slowing laser is sent in a direction counter-propagating with respect to the molecular beam. Field coils to remix the dark magnetic substates are placed between the cell and the detection region. }
\end{figure*}

The overall experimental setup is depicted in \rfig{fig:ExperimentalSetup}. CaF molecules are produced inside a 2\,K cell by ablation of a vacuum hot-pressed CaF$_{2}$ target with 7\,mJ, 4\,ns-long pulses from a frequency-doubled Nd:YAG laser at 532\,nm. The resulting CaF molecules are cooled by collisions with cold $^{4}$He buffer gas and extracted toward the detection region, which is 50\,cm downstream from the cell. The slowing lasers and the molecular beam counter-propagate. The laser light at 606\,nm and 628.5\,nm is generated by two ring dye lasers. An injection-locked external cavity diode laser (ECDL) is used as the \mysecond repump laser at 628\,nm. Four milliseconds after the molecules are generated, the main slowing laser is applied for a duration of 13 ms. The timing of the sequence is chosen to prevent molecules from being slowed too early to maximize the flux of molecules that reach the detection region in spite of their finite transverse velocity. The \myfirst and \mysecond repump lasers are applied from 2 ms to 20 ms after the ablation pulse. The intensities of the main, the \myfirst repump, and the \mysecond repump lasers are 260, 260, and 18 mW/cm$^{2}$, respectively. Dark magnetic substates generated by optical pumping precess about an applied transverse magnetic field (10 Gauss) and are returned to the optical cycle \cite{Shuman2010,Barry2012a}. 

A two-photon transition is used to detect the molecules. Laser light at 606 nm is sent orthogonally to the molecular beam in the detection region, exciting molecules from the $X\,^{2}\Sigma(v\Xlabel=0,N=1,J=1/2,F=1)$ state to the $A\,^{2}\Pi_{1/2}(v\Alabel=0,J=3/2,+)$ state. Velocity-selective detection is performed with 729.5 nm light intersecting the molecular beam at a 45 degree angle in the detection region (the center of our MOT chamber). This light excites molecules in the $A\,^{2}\Pi_{1/2}(v\Alabel=0,J=3/2,+)$ state to the $C\,^{2}\Pi_{1/2}(v\Clabel=0,J=1/2,-)$ state. UV photons are emitted at 331\,nm when the molecules decay from the $C\,^{2}\Pi_{1/2}$ state to the $X\,^{2}\Sigma$ state-manifold. These photons are detected by a photomultiplier tube. Due to the parity selection rule, molecules decay from the $C\,^{2}\Pi_{1/2}$ state to the $X\,^{2}\Sigma(v\Xlabel=0,N=0,2)$ states and are lost from the optical cycle. The 30 GHz rotational splitting between the $A\,^{2}\Pi_{1/2}(J=1/2,3/2)$ states makes the detection and slowing processes independent of each other.

\begin{figure*}
\begin{centering}
\includegraphics[width=5.5in]{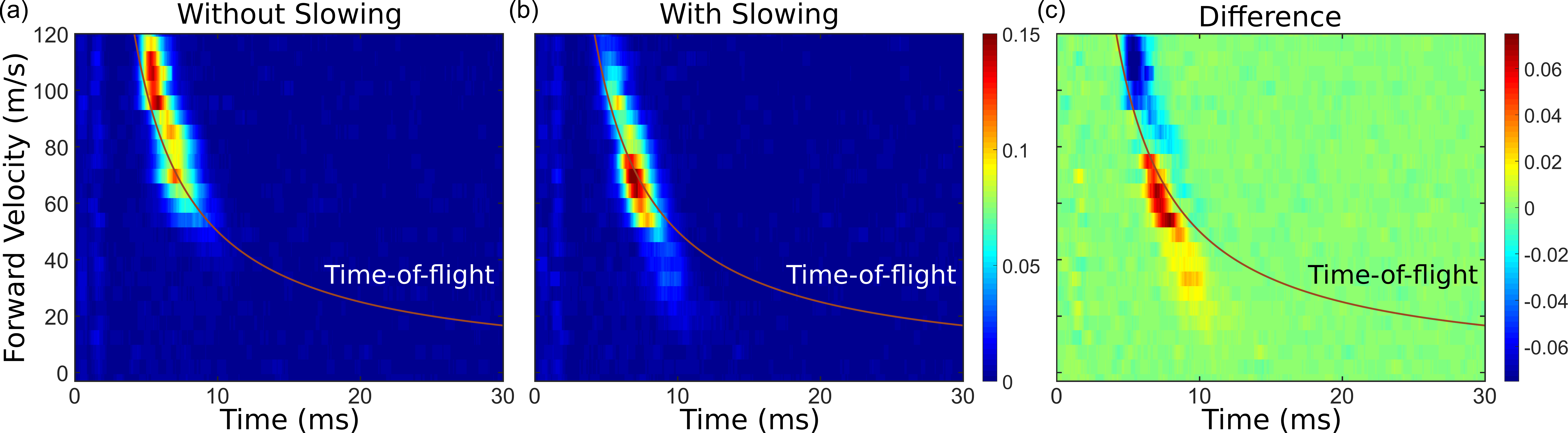}
\par\end{centering}

\protect\caption{\label{fig:SlowingScan}Laser slowing of CaF. The forward velocity of CaF is measured with and without the slowing ((a) and (b)). The color bar indicates the photomultiplier signal (in arbitrary units) which is proportional to the number of molecules. The difference between (a) and (b) is plotted in (c) and demonstrates the slowing effect: The velocity distribution is depleted of molecules with higher velocities (blue area), whereas molecules accumulate at lower velocities (red area) with the slowing laser on. Red lines in (a), (b), and (c) are time-of-flight lines as a guide to the eye. Each data set is the result of 100 averages.}

\end{figure*}

\begin{figure*}
\begin{centering}
\includegraphics[width=0.9\textwidth]{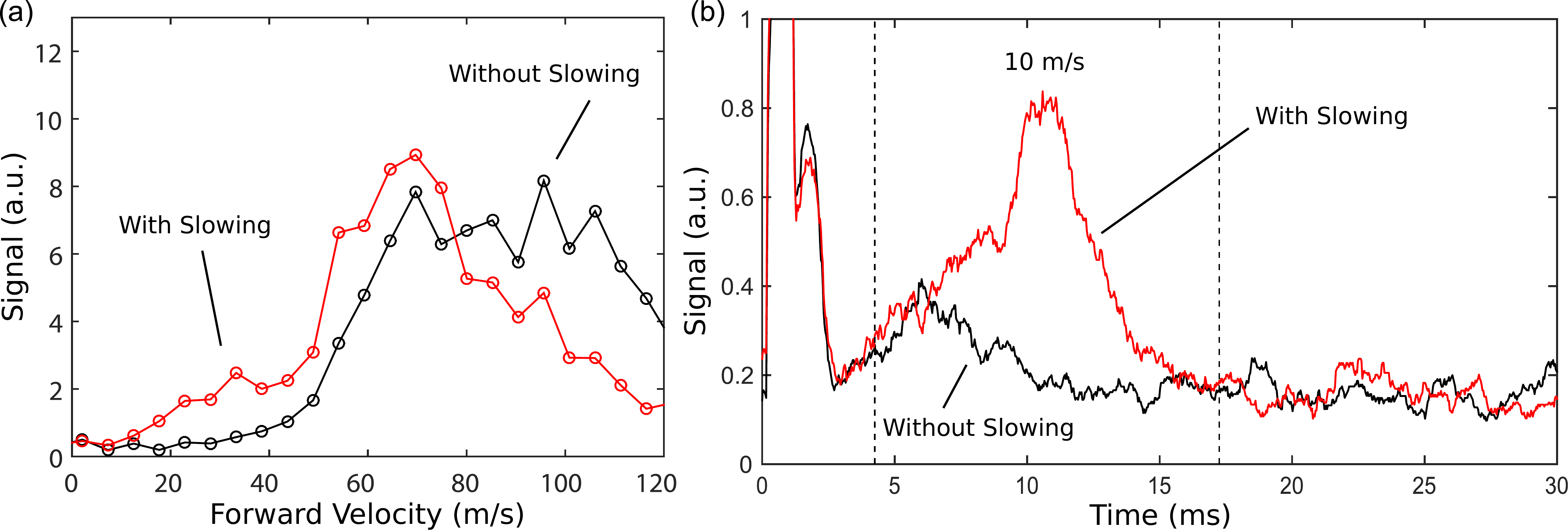}
\par\end{centering}

\protect\caption{\label{fig:SlowingTimetraces} (a) The signal integrated from 2 ms to 30 ms shows the overall velocity shift due to the slowing lasers. (b) CaF molecules slowed down to velocities between $10\pm 4$\,m/s. The signal at around 5\,ms without slowing laser is ablation laser-induced fluorescence from the experimental setup. The time window when the main slowing laser is applied is shown with the vertical dashed lines. This data set is the result of 3000 averages. The signal at 10\,m/s corresponds to approximately \moleculenumber molecules.}

\end{figure*}

\rfig{fig:SlowingScan} summarizes the results of the white-light slowing of CaF molecules. We compare the molecular laser-induced fluorescence signal with and without the main slowing laser while keeping the \myfirst and \mysecond repump lasers on for both cases. This removes the effect of optical pumping of the naturally populated excited vibrational states in the molecular beam. \rfig{fig:SlowingScan}(a) shows the CaF beam signal with only the repump lasers applied. The arrival time for each velocity agrees well with the time-of-flight hyperbola in the plot. When the slowing lasers are applied, molecules with velocities as low as 10\,m/s are observed (\rfig{fig:SlowingTimetraces}(b)). They arrive ahead of the predicted time from time-of-flight, indicating that they have been slowed down from a higher initial velocity. The time-integrated signal (\rfig{fig:SlowingTimetraces}(a)) shows that slowing lasers modify the velocity distribution of the beam: the number of molecules with speeds $>$\,80\,m/s has decreased and that of molecules with speeds $<$\,80\,m/s has increased. It should be noted that the white-light slowing process mainly shifts the velocity of the molecules and does not bunch them up at a final velocity due to the soft edge of our broadened light spectrum.  This together with the initial velocity distribution and transverse spreading determines the final velocity distribution.

The signal from the slowest molecules is shown in \rfig{fig:SlowingTimetraces}(b). Molecules with velocities between $10 \pm 4$\,m/s (which is near the expected capture velocity of a MOT for CaF) are observed. They arrive $\sim 10$\,ms after they leave the cell at 2\,ms, indicating that their initial longitudinal velocity was at least 50\,m/s. 
To within a factor of two, using the laser-induced fluorescence signal, we observe \moleculenumber molecules that are slowed down to near the MOT capture velocity. The primary contributions to the error are light collection efficiency of the optics, photomultiplier tube calibration and molecular beam fluctuations.

In summary, we have demonstrated laser slowing of CaF molecules from a two-stage buffer-gas beam. The slow initial velocity of our source allows slowing over a shorter distance ($\approx 20$\,cm); this, in turn, could lead to an overall increase in the number of molecules captured by a MOT. Once the molecular MOT is achieved, we plan to co-trap an atomic species to study atom-molecule collisions and the possibility of sympathetic cooling of CaF. A promising coolant atom for this endeavour is Li, where the ratio of elastic-to-inelastic scattering rates is predicted to be favorable \cite{Lim2015,Tscherbul2011}. We expect the co-loading of an atomic species to be straightforward since it has been demonstrated that an atomic MOT can be loaded directly from our buffer-gas source without additional laser slowing \cite{Hemmerling2014}. Generalization of such methods to molecules may pave the way to using ultracold molecules for probing new physics, such as the study of exotic phases of matter using the spin degree of freedom and long-range dipole-dipole interactions of polar molecules \cite{Micheli2006}. 

{{\it Acknowledgements.} We thank Benjamin Augenbraun for proofreading the manuscript. We acknowledge funding support from ARO and NSF.}


\vspace{1cm}

\bibliography{References}

\end{document}